\documentclass[twocolumn,prl,aps,superscriptaddress,floatfix,showpacs]{revtex4}
\usepackage{graphicx}

\begin{document}

\title{Phonon dispersion and lifetimes in $\rm MgB_2$ \\}

\author{Abhay Shukla}
\affiliation{Laboratoire de Min\'eralogie-Cristallographie, case 115, 4 Place Jussieu, 75252, Paris cedex 05, France}
\author{Matteo Calandra}
\affiliation{Laboratoire de Min\'eralogie-Cristallographie, case 115, 4 Place Jussieu, 75252, Paris cedex 05, France}
\author{Matteo d'Astuto}
\affiliation{European Synchrotron Radiation Facility, BP 220, F-38043 Grenoble cedex, France}
\author{Michele Lazzeri}
\affiliation{Laboratoire de Min\'eralogie-Cristallographie, case 115, 4 Place Jussieu, 75252, Paris cedex 05, France}
\author{Francesco Mauri}
\affiliation{Laboratoire de Min\'eralogie-Cristallographie, case 115, 4 Place Jussieu, 75252, Paris cedex 05, France}
\author{Christophe Bellin}
\affiliation{Laboratoire de Min\'eralogie-Cristallographie, case 115, 4 Place Jussieu, 75252, Paris cedex 05, France}
\author{Michael Krisch}
\affiliation{European Synchrotron Radiation Facility, BP 220, F-38043 Grenoble cedex, France}
\author{J. Karpinski}
\affiliation{Solid State Physics Laboratory, ETH, CH-8093 Z\"urich, Switzerland}
\author{S. M. Kazakov}
\affiliation{Solid State Physics Laboratory, ETH, CH-8093 Z\"urich, Switzerland}
\author{J. Jun}
\affiliation{Solid State Physics Laboratory, ETH, CH-8093 Z\"urich, Switzerland}
\author{D. Daghero}
\affiliation{INFM - Dipartimento di Fisica, Politecnico di Torino, C. Duca degli Abruzzi 24, 10129 Torino, Italy}
\author{K. Parlinski}
\affiliation{Institute of Nuclear Physics, ul. Radzikowskiego, 152, 31-342, Cracow, Poland.}

\date{\today}

\begin{abstract}
We measure phonon dispersion and linewidth in a single crystal of
$\rm MgB_2$ along the $\Gamma$-A, $\Gamma$-M and A-L directions using inelastic 
X-Ray scattering.
We use Density Functional Theory to compute the effect of both electron-phonon 
coupling and anharmonicity on the linewidth, 
obtaining excellent agreement with experiment.
Anomalous broadening of the $E_{2g}$ phonon mode is found all along $\Gamma$-A.
The dominant contribution to the linewidth is always the electron-phonon coupling.
\end{abstract}

\pacs{63.20.Dj, 63.20.Kr, 78.70.Ck, 71.15.Mb}

\maketitle

The discovery of 39 K superconductivity in $\rm MgB_2$ \cite{Nagamatsu} has led to
in-depth study of the material and a picture has emerged of a
phonon-mediated superconductor with multiple gaps \cite{Shulga,Liu,Choi},
moderate electron-phonon coupling (EPC) \cite{An,Kortus,Kong,Liu}, and
anharmonicity \cite{Liu,Yildirim,Choi,Boeri}.
However no measurements exist concerning either phonon dispersion or the evolution
of phonon lifetimes over the Brillouin Zone (BZ), due to the absence of
large single crystals. Neutron scattering on powder samples
\cite{Osborn,Yildirim} has been limited to the determination of phonon density of states.
Raman spectroscopy, which is restricted to the BZ center,
has shown that the optical mode with $E_{2g}$ symmetry,
corresponding to in-plane distortions of the B hexagons,
is strongly damped \cite{Bohnen,Goncharov,Postorino,Hlinka}.

Phonon damping can be caused by (i) EPC mediated phonon decay into electron-hole pairs \cite{Allen}, or
(ii) phonon-phonon interaction due to
anharmonicity \cite{Cardona}. The linewidth (the inverse of the lifetime) of a given
phonon is the sum of both contributions.
Direct determination of the  contribution of each phonon mode to EPC
\cite{Allen} from the measured linewidth is only possible if the anharmonic contribution
is negligible \cite{Grimvall} and is seemingly questionable for $\rm MgB_2$
where many calculations \cite{Yildirim,Choi,Liu} suggest strong anharmonic effects.

In this work we present the first measured phonon dispersion curves and
linewidths (where possible) along three directions in the BZ, 
$\Gamma$-A, $\Gamma$-M and A-L. 
We circumvented the problem of
sample size by using high resolution inelastic scattering of a focused and intense
X-Ray beam at the European Synchrotron Radiation Facility
(beamline ID28), a technique \cite{Ruf} successfully used in single crystalline
samples and in particular for the measurement of high energy optical 
modes \cite{Dastuto}.
To understand the mechanisms governing the measured phenomena,
we calculated phonon dispersion, the contributions of EPC and 
anharmonicity to the linewidth and the structure factors 
 using Density Functional Theory (DFT).

Small single crystals of $\rm MgB_2$, suitable for inelastic X-ray scattering experiments have recently
become available. The crystal used in our experiment was grown at a pressure of
30-35 kbar.
A mixture of Mg and B was put into a BN container in a cubic anvil device.
The temperature was increased during one hour up to
$\rm 1700-1800^{\circ}$C, kept stable for 1-3 hours and decreased
during 1-2 hours. As a result plate-like $\rm MgB_2$ crystals were formed of which we
used a sample of about 400 x 470 x 40 $\mu m^3, $ with a measured
in-plane mosaicity of $\rm 0.007^{\circ}$.
The beam incident on the sample was obtained from a high-resolution Silicon
backscattering monochromator using the (8 8 8) reflection at an incident energy of
15.816 keV. The X-ray beam was focused onto the sample by a toroidal mirror
into a spot of 270 x 90 $\mu m^2$(horizontal $\times$ vertical), 
full width at half maximum
(FWHM). Slits before the sample further limited
the vertical beam size to 30 $\mu m$. The scattered photons were analyzed in
energy by five spherical silicon crystal analyzers operating at the same reflection
order and mounted in pseudo Rowland circle geometry.
The total energy resolution was 6.1 meV FWHM, as determined by a fit to a
Lorentzian  lineshape. The momentum transfer {\bf Q} was selected by rotating the
7 m long analyzer arm around the sample position, in the horizontal plane, which
also contained the linear x-ray polarization vector of the incident beam.
The momentum resolution was set to $\rm 0.04 {\AA}^{-1}$ in the horizontal direction
and $\rm 0.07 {\AA}^{-1}$ in the vertical direction.
The following measurements were performed at a temperature of 300K:
i) {\bf Q}=(1 2  $\xi$), in almost transverse configuration along the $\Gamma$-A direction,
i.e.  with ${\bf Q \cdot q} \approx 0$, {\bf q}=(0 0 $\xi$) being the phonon wavevector,
ii) {\bf Q}=(1-$\delta$ 1+$\xi$ 0), while nearly following the $\Gamma$-M direction ($ 0 \leq \delta \leq 0.05$).
iii) {\bf Q}=(1-$\delta_1$ 2+$\xi$ 0.5+$\delta_2$), while nearly following the A-L direction ($0 \leq \delta_1,\delta_2 \leq 0.04$).
The choice of BZ points measured
was dependent on a series of conditions including the need to optimize the structure factors,
single ($\Gamma$-A) or multi-analyzer ($\Gamma$-M, A-L) measurement mode and spectrometer and time limitations.

\begin{figure}[ht]
\includegraphics[width=8.5cm]{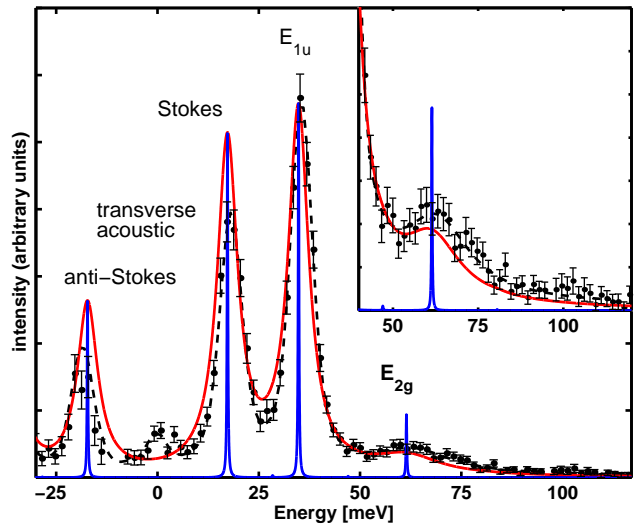}
\caption{(color online) Energy loss scan in almost transverse
geometry measured at {\bf Q}=(1 2 0.3) corresponding to 0.6 $\Gamma$-A.
The data, normalized to the incident flux, are shown with the least-squares fit 
(dashed line) and the ab-initio spectrum  with and 
without broadening due to experiment and electron phonon
coupling (solid lines).
The broad peak corresponds to the damped $E_{2g}$ mode and is
shown in greater detail in the inset.The peak at zero is due to diffuse scattering.
} \label{fig1}
\end{figure}

In Fig.\ \ref{fig1} we show data taken at the 0.6 $\Gamma$-A point in the BZ.
The acoustic mode as well as the lower energy
optical mode ($E_{1u}$) are visible as resolution-limited peaks.
Most importantly, a broad peak is observed at higher energy loss,
corresponding to the $E_{2g}$ mode. We performed least square fits to sums of 
Lorentzian functions with FWHM corresponding to the experimental resolution for the
resolution limited peaks and a free parameter for the strongly damped phonon. 
These yield the dispersion as well as the linewidth variation over the BZ.
Despite statistical limitations (3-6 counts per minute on this peak
along $\Gamma$-A) and tails of the peaks from the
stronger, low energy phonons, the peak energy as well as the linewidth can be
estimated with reasonable confidence.

\begin{figure}[ht]
\includegraphics[width=8.5cm]{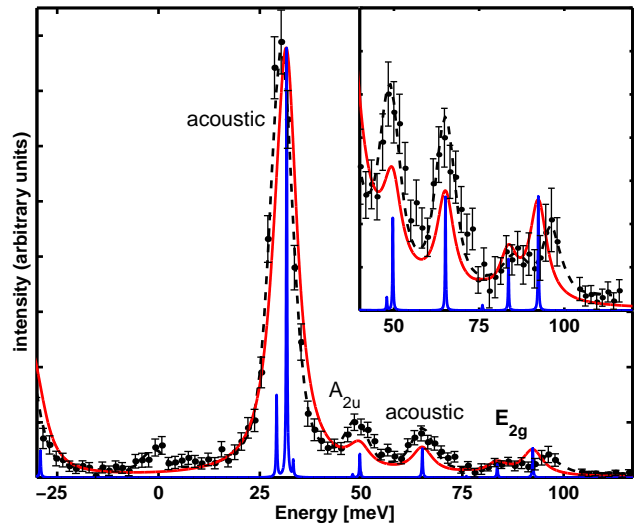}
\caption{(color online) Energy loss scan, measured at {\bf Q}=(0.97 2.29 0.54)
corresponding closely to 0.58 A-L, with the least-squares fit and the
ab-initio spectrum calculated at 0.58 A-L.} \label{fig2}
\end{figure}

Fig.\ \ref{fig2} shows a similar energy loss scan
at (0.97 2.29 0.54) close to the 0.58A-L point.
A strong acoustic mode is seen at 30 meV.
The peak at 50 meV corresponds to the  $A_{2u}$ branch and the one at 65 meV 
to another acoustic branch. Finally two resolved features are seen at 85 
and 97 meV. These are the two $E_{2g}$ modes which in this region of reciprocal 
space are well separated from other modes. Though, given the statistics, it would 
be hazardous to estimate a linewidth,  the 
comparison between the experimental and ab-initio spectra 
suggests that the linewidths of both the $E_{2g}$ modes are resolution
limited and so the damping is much less than that along  
$\Gamma$-A.
As for the measurement nearly along $\Gamma$-M, the structure factor for the 
optical modes strengthens only near the zone boundary. 
At the point measured nearest to M ((1.05 1.45 0), not shown) the $E_{2g}$ and 
$E_{1u}$ modes are comparable in intensity but only separated by about 1 meV 
according to our calculation and we do measure a single peak only somewhat 
broader (FWHM $\approx 10meV$) than the experimental resolution, 
indicating reduced $E_{2g}$ linewidth. The proximity of the $E_{1u}$ mode however 
prevents a firm conclusion in this regard. We mention that the calculated 
structure factors and energies show excellent quantitative agreement with our 
measured data of which we have shown only two examples.

\begin{figure}[ht]
\includegraphics[width=8.5cm]{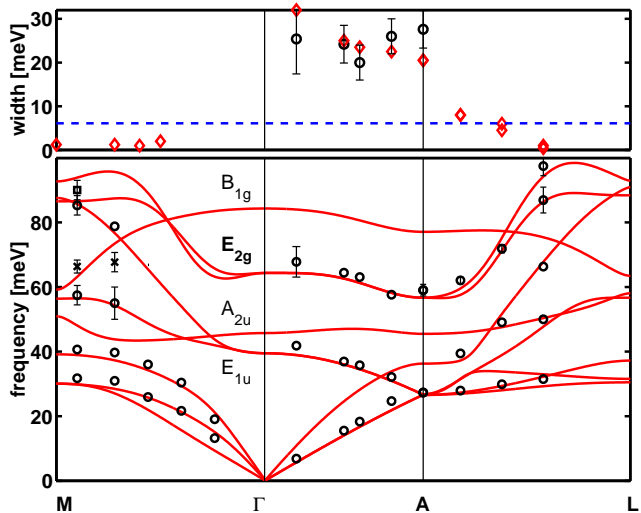}
\caption{(color online) Bottom: Experimental (circles) and theoretical phonon dispersion
 (solid line) in $\rm MgB_2$ along $\Gamma$-A and $\Gamma$-M and A-L. 
In the region near the M point, the probable detection of the $E_{2g}$
mode is indicated with a square symbol (see text). The crosses indicate a parasite 
signal of unknown origin.
Top: Intrinsic linewidth of the $E_{2g}$ mode. The experimental linewidth (circles)
 is large along $\Gamma$-A and below the experimental resolution (dashed line)
 near L and M. The theoretical result (diamonds) for the
electron-phonon coupling contribution to the linewidth is also shown.
Along A-L and $\Gamma$-M where the $E_{2g}$ mode
is non-degenerate both theoretical values are shown, when different.
$E_{2g}$ linewidth decreases progressively from A to L and 
experimentally the two branches are
resolved for the point nearest to L.
}
\label{fig3}
\end{figure}

Similar analysis was done for several points along the three directions
in order to experimentally determine the phonon dispersion and the linewidths. The 
difference in calculated phonon energies between the measured points and 
corresponding points exactly
along $\Gamma$-M and A-L ($\delta,\delta_1,\delta_2 = 0 $) is less than half a meV in 
all cases. We can thus compare the experimental phonon dispersion with 
the theoretical calculation along the high symmetry lines,
as shown in the bottom panel of fig. \ref{fig3} (circles).

The measured intrinsic linewidth of the $E_{2g}$ branch, 
shown in the top panel of fig.\ \ref{fig3}, is
strongly anisotropic in the BZ. Along $\Gamma$-A it is particularly large
(ranging from 20 to 28 meV), while near L and probably near M it is  below
the experimental resolution. 

Electronic structure calculations \cite{PWSCF} were performed using
DFT in the generalized gradient approximation
\cite{PBE}. We used norm conserving pseudo-potentials \cite{Troullier}.
For $\rm Mg$ we used non-linear core corrections \cite{NLCC}
and we treated the $2s$, $2p$ levels as core states.
The wave-functions were expanded in plane waves using a $35$ Ry cutoff.
The calculations were performed with the experimental
crystal structure, namely $a=3.083 \AA$ and $c/a=1.142$.
The harmonic phonon frequencies were computed in
the linear response \cite{DegironcPRB}.
We used a $16\times 16\times 16$ Monkhorst-Pack grid
for the electronic BZ integration and first order Hermite-Gaussian smearing \cite{Degironcsmear}
of $0.025$ Ry.
The dynamical matrix at a given point of the BZ was obtained
from a Fourier interpolation of the dynamical matrices computed
on a $6\times 6\times 4$ phonon mesh. The resulting phonon frequencies
are shown in fig. \ref{fig3} and are in good agreement with a recent calculation
\cite{Bohnen}.
The agreement with experiment is remarkable.

The contribution to the FWHM linewidth $\gamma_{{\bf q}\nu}$ at momentum ${\bf q}$
for the $\nu$ phonon mode  due to the
electron-phonon interaction can be written as \cite{Allen}:
\begin{equation}\label{eq:eplw}
\gamma_{{\bf q}\nu} = \frac{4\pi \omega_{{\bf q}}}{N_{k}} \sum_{{\bf k},n,m} |g_{{\bf k}n,{\bf k+q}m}^{\nu}|^2 \delta(\varepsilon_{{\bf k}n}) \delta(\varepsilon_{{\bf k+q}m})
\end{equation}
where the sum is extended over the BZ, $N_k$ is the number of $k$-points in the sum,
and $\varepsilon_{{\bf k}n}$ are the energy bands measured with
respect to the Fermi level at point ${\bf k}$.
The matrix element is
$g_{{\bf k}n,{\bf k+q}m}^{\nu}= \langle {\bf k}n|\delta V/\delta u_{{\bf q}\nu} |{\bf k+q} m\rangle /\sqrt{2 \omega_{{\bf q}\nu}}$,
where $u_{{\bf q}\nu}$ is the amplitude of the displacement of the phonon $\nu$
of wavevector ${\bf q}$,
$\omega_{{\bf q}\nu}$ is the phonon frequency and $V$ is the Kohn-Sham
potential.

In the calculations we used $N_k=30^3$ inequivalent $k$-points and, in
Eqs.\ (\ref{eq:eplw}) and (\ref{bordello}), we substituted the $\delta$
functions with Gaussians.
The electron phonon coupling $\lambda_{{\bf q}\nu}$  is obtained from the linewidth
\cite{Allen} as:
\begin{equation}\label{lambda}
\lambda_{{\bf q}\nu}=\frac{\gamma_{{\bf q}\nu}}{2\pi N(0) \omega_{{\bf q}\nu}^2},
\end{equation}
$N(0)=0.354$ states/(${\rm MgB_2}$ eV spin) being the density of states at the
Fermi level.

The second contribution to the linewidth is given by the anharmonicity
in the crystal potential. At lowest order for the mode $\nu$
of a zone center phonon the  FWHM linewidth is \cite{Cardona,Debernardi,Lang}:
\begin{equation}\label{bordello}
\Gamma_{{\bf 0}\nu}= \frac{\pi \hbar}{8 N_q} \sum_{{\bf q},\mu \eta}
|\frac{\partial^{3} E}{\partial u_{{\bf 0}\nu} \partial u_{{\bf q}\mu}\partial u_{{\bf -q}\eta}}|^2
\frac{I_{{\bf q}\mu\eta\nu}^{\rm D}+I_{{\bf q}\mu\eta\nu}^{\rm A}}{\omega_{{\bf 0}\nu}\omega_{{\bf q}\mu} \omega_{{\bf q}\eta}}
\end{equation}
$E$ being the total energy, $n_{{\bf q}\mu}$ the Bose occupation for mode $\mu$ at
wavevector ${\bf q}$,  $I^{\rm D}_{{\bf q}\mu\eta\nu}=(n_{{\bf q}\mu}+n_{{\bf q}\eta}+1)\delta\left(\omega_{{\bf 0}\nu}-\omega_{{\bf q}\mu}-\omega_{{\bf q}\eta}\right)$ describes
the decay in the two phonons $\mu$ and $\eta$,  and
$I^{\rm A}_{{\bf q}\mu\eta\nu}=2(n_{{\bf q}\mu}-n_{{\bf q}\eta})\delta\left(\omega_{{\bf 0}\nu}-\omega_{{\bf q}\mu}+\omega_{{\bf q}\eta}\right)$
describes the $\eta$-phonon absorption and the $\mu$-phonon emission.

We computed the anharmonic linewidth  at
the high-symmetry points $\Gamma$, A, M.
For the calculation at A we
consider a $1\times 1 \times 2$ supercell  with 6 atoms,
while for the M point we use a
$2\times 2\times 1$ cell with 12 atoms.
The third order matrices
were evaluated using linear response theory and the $2n+1$ theorem
for metals \cite{Lazzeri}. The anharmonic contribution was evaluated
at $0K$ and $300K$.

At $\Gamma$ the anharmonic linewidth is largest for the $E_{2g}$ mode and
equal to  $0.16$ meV at $T=0K$ and $1.21$ meV at $T=300K$.
Both the values are negligible if compared with
the experimental Raman linewidth of roughly $40$ meV \cite{Postorino}, suggesting
that the main source of broadening is the electron-phonon interaction.

The results of the calculation of the two contributions to the linewidth at
A and M are shown in Table \ref{tabella}.
At the A point the linewidth of the $E_{2g}$ mode
due to electron-phonon scattering is very large
while the anharmonic contribution is more than an order of magnitude smaller.
For sizable values of the EPC and at low temperature the anharmonicity
is negligible showing that a measurement of the linewidth at ${\bf q}$
for the mode $\nu$ is essentially equivalent to determining $\lambda_{{\bf q}\nu}$.
This is unexpected since earlier theoretical work \cite{Yildirim,Choi,Liu}
estimates anharmonicity to be important for a different 
but related quantity, the frequency shift.
\begin{table}
\begin{ruledtabular}
\begin{tabular}{cccccccc}
  \multicolumn{4}{c}{M}& \multicolumn{4}{c}{A} \\
 $\Gamma_{\nu}^{0}$ & $\Gamma_{\nu}^{300}$  & $\gamma_{\nu}$ & $\lambda_{\nu}$ &  $\Gamma_{\nu}^{0}$ & $\Gamma_{\nu}^{300}$  & $\gamma_{\nu}$ & $\lambda_{\nu}$ \\ \hline
0.00 &   0.12 &   0.00 &   0.00    &   0.00 &   0.17 &   0.00 &   0.00 \\
0.00 &   0.15 &   0.01 &   0.01    &   0.00 &   0.17 &   0.00 &   0.00 \\
0.02 &   0.12 &   0.06 &   0.02 &   0.00 &   0.63 &   0.08 &   0.05 \\
0.12 &   0.48 &   1.13 &   0.20 &   0.00 &   0.63 &   0.08 &   0.05 \\
0.06 &   0.25 &   0.00 &   0.00 &   0.02 &   0.22 &   0.84 &   0.28 \\
0.07 &   0.33 &   2.34 &   0.30 &   0.02 &   0.20 &   0.08 &   0.02 \\
{\bf 0.26} &  {\bf 0.42} & {\bf  1.06} &  {\bf 0.06} & {\bf  0.10} & {\bf  2.13} & {\bf  20.35} & {\bf  2.83} \\
0.45 &   0.69 &   1.21 &   0.07 &  {\bf 0.10} & {\bf  2.13} & {\bf 20.35} & {\bf  2.83} \\
{\bf 0.47} &   {\bf 0.72} &   {\bf 0.08} &  {\bf 0.00} &   0.13 &   0.23 &   0.05 &   0.00 \\
\end{tabular}
\end{ruledtabular}
\caption{Calculated linewidths (meV) due to anharmonicity at
$0$K ($ \Gamma_{\nu}^{0}$), $300$K ($\Gamma_{\nu}^{300}$)
and electron-phonon interaction $\gamma_{\nu}$ at M and A
for all modes. Phonon frequencies increase from top to bottom.
$\lambda_{\nu}$ is the electron-phonon coupling. $E_{2g}$ modes in boldface. }
\label{tabella}
\end{table}

\begin{table}
\begin{ruledtabular}
\begin{tabular}{lccccc}
{\bf q} & 0.2 $\Gamma$-A & 0.5 $\Gamma$-A & 0.6 $\Gamma$-A & 0.8 $\Gamma$-A & 1.0 $\Gamma$-A \\
$\lambda_{{\bf q}\nu}^{\rm expt.}$ & 2.5$\pm$1.1 &  2.6$\pm$0.6 & 2.3$\pm$0.5 & 3.6$\pm$0.7 & 3.6$\pm$0.8 \\
$\lambda_{{\bf q}\nu}^{\rm theo.}$ & 3.32 & 2.80 & 2.77 &  3.12 &   2.83 \\
\end{tabular}
\end{ruledtabular}
\caption{Experimental and theoretical $\lambda_{{\bf q}\nu}$ of each of the two degenerate $E_{2g}$ modes along $\Gamma$-A.}
\label{tabella2}
\end{table}

Since the anharmonic contribution is negligible at $\Gamma$, A, M, 
and computationally demanding, we only evaluated the EPC linewidth for the other
points along the three directions. 
Using the calculated phonon frequencies, 
displacements and linewidths we compute the structure 
factors for one phonon processes  using the X-ray form 
factors, obtaining good agreement with experiment as shown in 
Figs.\ \ref{fig1},\ref{fig2}.  
In the top panel of Fig.\ \ref{fig3} we show the theoretical results
for the $E_{2g}$ branch. The result is consistent with the experimental value
wherever the $E_{2g}$ branch is clearly visible namely along $\Gamma$-A and near 
the L point. 
Finally, the anharmonicity being negligible, we used Eq. (\ref{lambda}) to
extract $\lambda_{{\bf q}\nu}$ of the $E_{2g}$
mode along $\Gamma$-A using the measured linewidths and frequencies together with
the calculated electronic density of states.
In Table \ref{tabella2} the experimental values
are compared with the theoretical predictions. 
The anomalously large EPC along $\Gamma$-A is due to
the nesting factor of the B bonding $p_{x,y}$ Fermi surfaces, 
which are concentric cylinders centered on $\Gamma$-A \cite{An,Kortus}. 
The $E_{2g}$ modes, which modify the B-B distances, are the
only ones with a sizeable matrix element, 
$g_{{\bf k}n,{\bf k+q}m}^{\nu}$, between electrons on these
surfaces.

In conclusion, we have measured phonon dispersion and linewidths in a sub-mm sized
$\rm MgB_2$ crystal with inelastic X-ray scattering confirming the power and
versatility of this technique. Both acoustic and optical modes are detected
and we find that the $E_{2g}$ mode is anomalously broadened
along $\Gamma$-A but that this broadening is not generalized over the
Brillouin Zone.
Our Density Functional Theory calculations of the dispersion and linewidth 
are in excellent agreement with experiment.
They show that the dominant contribution to the broadening for all modes 
is the electron-phonon coupling, the anharmonic contribution being much smaller.
Thus phonon linewidth in $\rm MgB_2$ is a direct measure of electron-phonon 
coupling and could, with the availability of larger samples, be measured for all modes over
the whole Brillouin zone so as to extract the anisotropic Eliashberg coupling function. 

We acknowledge illuminating discussions with R. S. Gonnelli,  P. Giannozzi,
M. Xu,  and F. Sette.
The calculations were
performed at the IDRIS supercomputing center. M.C. was supported by a
Marie Curie Fellowship of the European Commission, contract No. IHP-HPMF-CT-2001-01185.

\end{document}